\documentclass[
twocolumn,       
    aps, prd,          
    superscriptaddress, 
    secnumarabic,      
nofootinbib,     
    amsmath,           
    amssymb            
]{revtex4-1}
\usepackage{graphicx}      
\usepackage{wrapfig}       
\usepackage{dcolumn}       
\graphicspath{{./}{figures_model/}{figures/}{other/}{figures_new_initial_condition/}}
\usepackage{amsmath}                
\usepackage{amssymb}                
\usepackage{eqnarray}               
\usepackage{bm}

\usepackage{pifont}                 
\usepackage{wasysym}                
\usepackage{textcomp}               

\usepackage[dvipsnames]{xcolor}     
\usepackage{soul,ulem}              
\soulregister\cite7
\soulregister\ref7 
\soulregister\eqref7                

\usepackage{cancel}                 
\usepackage{comment}                
\usepackage{marginnote}             
\usepackage[colorinlistoftodos,
  textsize=tiny]{todonotes}         

\usepackage[colorlinks=true,linkcolor=blue]{hyperref}%
\usepackage{cleveref}               
\hypersetup{
    citecolor=darkgreen
}

\usepackage{algorithm}              

\usepackage[ddmmyyyy]{datetime}     
\usepackage{acro}          
\DeclareAcronym{FLRW}{
  short = FLRW ,
  long = Friedmann-Lema{\^i}tre-Robertson-Walker
}
\DeclareAcronym{GR}{
  short = GR ,
  long = general relativity
}
\DeclareAcronym{BAO}{
  short = BAO ,
  long = Baryon acoustic oscillations
}
\DeclareAcronym{CMB}{
  short = CMB ,
  long = Cosmic microwave background
}
\DeclareAcronym{CMBR}{
  short = CMBR ,
  long = cosmic microwave background radiation
}

\DeclareAcronym{GW}{
  short = GW ,
  long = gravitational wave
}

\DeclareAcronym{RHS}{
  short = RHS ,
  long = right hand side
}

\DeclareAcronym{CDM}{
  short = CDM ,
  long = cold dark matter 
}

\DeclareAcronym{STEGR}{
  short = STEGR ,
  long = Symmetric Teleparallel Equivalent of GR
}

\DeclareAcronym{TEGR}{
  short = TEGR ,
  long = Teleparallel Equivalent of GR
}

\DeclareAcronym{DE}{
  short = DE ,
  long = dark energy
}

\DeclareAcronym{MG}{
  short = MG ,
  long = modified gravity
}

\definecolor{mypurple}{RGB}{135,10,190} 
\definecolor{mygreen}{RGB}{0,130,0} 
\definecolor{softgreen}{RGB}{52, 168, 83} 
\definecolor{emerald}{RGB}{80,200,120}
\definecolor{forestgreen}{RGB}{34,139,34}  
\definecolor{mediumseagreen}{RGB}{60,179,113} 
\definecolor{darkgreen}{RGB}{0, 128, 0}

\newcommand{\beq}{\begin{equation}}
\newcommand{\eeq}{\end{equation}}
\newcommand{\be}{\begin{equation}}
\newcommand{\ee}{\end{equation}}
\newcommand{\bea}{\begin{eqnarray}}
\newcommand{\eea}{\end{eqnarray}}
\newcommand{\eqeqref}{Eq. \eqref}

\newcommand{\GN}{\ensuremath{G_{\rm N}}}

\newcommand{\Panp}{Pan$^{+}$}

\newcommand{\diff}{{\rm d}}

\begin{document}

\title{Dynamical Dark Energy or Modified Gravity? Signatures in Gravitational Wave Propagation}

\author{Purnendu Karmakar}
 \email{purnendu.physics@gmail.com}
 \affiliation{Institute of Physics, University of Tartu, W. Ostwaldi 1, 50411 Tartu, Estonia}

\author{Sandeep Haridasu}
 \email{sharidas@sissa.it}
 \affiliation{SISSA, Via Bonomea 265, 34136 Trieste, Italy}
\affiliation{IFPU - Institute for fundamental physics of the Universe, Via Beirut 2, 34014 Trieste, Italy}
 \affiliation{INFN-Sezione di Trieste, via Valerio 2, 34127 Trieste, Italy}

\begin{abstract}

We discuss the implications of the most recent DESI Baryon Acoustic Oscillations (BAO) DR2 and DESyr5 compilation of supernovae (SNe) datasets for modified gravity focusing on non-metricity based $f(Q)$ theory, by employing a `model-independent' approach. We reconstruct of dark energy density which is then extended to estimate the perturbation-level quantities, sourced by the modified gravity, namely the effective gravitational coupling, $\mu$, and the amplitude damping parameter of gravitational wave propagation, $\nu$. In light of the remarkable hints for a dynamical dark energy emerging from the analysis of background cosmological data, we discuss the possibility to distinguish between dark energy and modified gravity scenarios from their scalar and tensor  perturbation-level signatures. We contrast our findings between the older Pantheon+ and the newer DESyr5 compilations of SNe datasets, which predict a significant distinct signature in the damping parameter of the gravitational wave propagation amplitude. On the other hand, the prediction for the effective gravitational coupling remains insensitive to the choice of the SNe datasets. Our results provide a strong case for the study of gravitational wave propagation in modified gravity theories, in light of the new generation of gravitational wave detectors such as LISA, Einstein Telescope and Cosmic Explorer. 

\end{abstract}

\date{\today} 
\maketitle

\section{Introduction}
\label{sec:Introduction}
Alternative geometric reformulations of \ac{GR} \cite{BeltranJimenez:2019esp}, particularly those based on the affine connection consisting torsion and non-metricity, have attracted increasing interest in recent years. These approaches provide new perspectives on the nature of gravity comes from geometrical modification. The existence of equivalent formulations of \ac{GR}, such as the teleparallel equivalent (\ac{TEGR}) based on liner torsion scalar, and the symmetric teleparallel equivalent (\ac{STEGR}) based on linear non-metricity scalar \cite{Nester:1998mp, BeltranJimenez:2019tjy}, underscores the physical relevance of these extensions and motivates their systematic exploration in the context of modified gravity theories \cite{Bahamonde:2021gfp}.

One could extend these geometrical models further as a function of scalar curvature, called $f(R)$ model \cite{DeFelice:2010aj}, torsion called $f(T)$ model \cite{Cai:2015emx}, and non-metricity, called $f(T)$. 
Such modifications of gravity allow us explore the more phenomenonlogy of the universe both in the background and perturbation. In this article we take $f(Q)$ as an example, since it recently received decent attentions \cite{Barros:2020bgg, Ayuso:2020dcu, Bajardi:2020fxh, Frusciante:2021sio, Atayde:2021pgb}. 

A model-independent approach to reconstruct $f(Q)$ from observational data is therefore timely. Recent high-precision datasets, such as most recent DESI BAO DRII and Pantheon+, provide an opportunity to infer the shape of the function and the corresponding action, offering added value to the theoretical community.
In this work, we address phenomenological testing of $f(Q)$ models and reconstruct their action from current background and perturbation data using symbolic regression and rank the models preferred by the data. Our analysis also highlights potential observational distinctions between dark energy and modified gravity within this class of theories.

A wide range of functional forms for $f(Q)$ gravity have been investigated in the literature. Examples include exponential models \cite{Anagnostopoulos:2022gej, Yang:2024tkw, Gadbail:2024een, Su:2024avk, Gadbail:2024rpp, Mhamdi:2024xqd, Vasquez:2025hrz, Ayuso:2025vkc, Roy:2025nde, Boiza:2025xpn}, 
power-law models \cite{BeltranJimenez:2019tme, Anagnostopoulos:2022gej, Frusciante:2021sio, Bhardwaj:2024mop, Gadbail:2024een, Sahlu:2024pxk, Enkhili:2024dil, Gadbail:2024rpp, Kolhatkar:2024oyy, Sakr:2024eee, Wang:2024eai, Mhamdi:2024xqd, Yang:2024tkw, Ayuso:2025vkc, Paliathanasis:2025hjw, Roy:2025nde, DAgostino:2022tdk}, 
hyperbolic tangent forms \cite{Anagnostopoulos:2022gej, Su:2024avk}, 
logarithmic extensions \cite{Najera:2023wcw, Anagnostopoulos:2021ydo, Anagnostopoulos:2022gej, Goncalves:2024sem, Gadbail:2024een, Sakr:2024eee, Roy:2025nde, Karmakar:2025iba} 
models, and among others \cite{Jarv:2023sbp, Bahamonde:2022zgj, Heisenberg:2023lru}. In addition, several efforts have been devoted to reconstructing the dynamical connection \cite{Kar:2021juu, Capozziello:2022wgl, Gadbail:2022jco, Gadbail:2023klq, Naik:2023ykt, Mahmood:2023mac, Gadbail:2023mvu, Kaczmarek:2024yju, Gadbail:2024rpp, Kaczmarek:2024quk, Gadbail:2024lek, ElOuardi:2025okl, Gadbail:2024rpp, Yang:2024tkw, Yadav:2024vmt, Saha:2024gmk, Esposito:2021ect, Kang:2021osc, Nojiri:2024zab}, while other works adopt a designer approach \cite{Albuquerque:2022eac}, and addressing cosmological tensions \cite{Sakr:2024eee}. 

In this work, we explore the reconstruction of viable $F(Q)$ cosmological models directly from data, using background probes such as BAO and Pantheon+, along with perturbation-level observables including the effective gravitational constant $(\mu)$ and gravitational-wave propagation modification in amplitude damping $(\nu)$. While traditional approaches often rely on parameterizations of the dark energy equation of state $w(z)$, our method instead parameterizes the dark energy density and reconstructs the underlying $F(Q)$ function. This choice is more natural in the context of modified gravity, where the dynamics of the background and perturbations are tied to the fundamental geometric Lagrangian rather than an assumed equation of state of an energy density component.

At present, the available data lack the precision necessary to decisively distinguish modified gravity from dark energy at the level of perturbations. In particular, current measurements of growth $(f\sigma_8)$ and gravitational wave modifications are not yet sufficient to break the degeneracy between competing scenarios. Nevertheless, our reconstruction delivers model-independent predictions for $\mu$ and $\nu$, which can serve as benchmarks for upcoming surveys such as DESI, Euclid, SKA, and next-generation gravitational wave detectors LISA, DECIGO. These forecasts highlight the potential of future data to test $F(Q)$ gravity at both the background and perturbation levels, making our method a forward-looking tool to probe the nature of cosmic acceleration.

The issue of theoretical stability has also been discussed in detail \cite{Hu:2022anq, DAmbrosio:2023asf, Tomonari:2023wcs, Guzman:2023oyl, Gomes:2023tur, Bello-Morales:2024vqk}. In the present work, however, our emphasis lies on the phenomenological implications of $f(Q)$ cosmology, particularly through reconstruction techniques, which are versatile and can be applied across different classes of models.

The paper is organized as follows. In Section \ref{sec:fQgravity}, we briefly review the $f(Q)$ gravity and present the background and perturbation equations. In Section \ref{sec:modeling}, we describe the datasets and the methodology employed in this work. The results are presented in Section \ref{sec:resutls}, followed by a discussion and conclusion in Section \ref{sec:conclusions}. 

\section{f(Q) gravity}
\label{sec:fQgravity}

We start with the action of $f(Q)$ gravity framework,   
 \begin{equation}   
 \label{ac:fq:grav}
\mathcal{S}  =   \int {\diff^4} x \sqrt{-g} \Big[ - \frac{1}{2\kappa^2} f(Q) + \mathcal L_M (\rm{CDM, b, rad})\Big]\,.
 \end{equation}
with $\kappa^2 = 8\pi G_N$, $c=1$ and Newton's constant, $G$. 
The matter Lagrangian $\mathcal L_M (\rm{CDM, b, rad})$ represents the contributions from relativistic matter (photons, neglecting neutrinos) and non-relativistic matter (\ac{CDM} and baryons)
$f(Q)$ is an arbitrary function of the non-metricity scalar ($Q$). 
The non metricity scalar is defined as the contraction of the non-metricity tensor $Q_{\alpha\mu\nu}$, with its conjugate, $P^{\alpha\mu\nu}$, 
\beq
Q = -Q_{\alpha\mu\nu}P^{\alpha\mu\nu}\,.
\eeq
The non-metricity tensor is given as, $Q_{\alpha\mu\nu}=\nabla_\alpha g_{\mu\nu}$, which by definition incompatible with metric. 
The non-metricity conjugate is constructed combining its two independent traces $Q_\alpha=g^{\mu\nu}Q_{\alpha\mu\nu}$ and $\tilde{Q}_\alpha=g^{\mu\nu}Q_{\mu\alpha\nu}$ 
\begin{equation}
P^\alpha{}_{\mu\nu} = -\frac{1}{2}L^\alpha{}_{\mu\nu} + \frac{1}{4}\left( Q^\alpha - \tilde{Q}^\alpha \right) g_{\mu\nu} - 
\frac{1}{4}\delta^\alpha_{(\mu}Q_{\nu)}\,.
\end{equation}

The non-metricity conjugate tensor satisfies the variation, 
\beq
P^{\alpha\mu\nu}=-\frac12\frac{\partial Q}{\partial Q_{\alpha\mu\nu}}.
\eeq

We assume a homogeneous, isotropic and spatially flat \ac{FLRW} \cite{Green:2014aga} background, whose with metric, $\diff s^2 = -\diff t^2 + a^2(t)\delta_{ij}\diff x^i \diff x^j$. This yields Hubble expansion, $H=\dot{a}/a$, where $a(t)$ is the scale factor and its derivative are in cosmic time, $t$. The time evolution of the Hubble parameter satisfies $\frac{\ddot a}{a} = \dot H + H^2$. 

Three sets of coordinate choices for the connections has been discussed in the past lituratures. From here onward, we adopt the simplest coordinate choice, $Q=6H^2$ \cite{Gomes:2023tur}, that results in the disformation equal to negative Levi-Civita connection for the non in the absence of torsion. The coincident gauge can also be imposed together with the coordinate choice, though not part of this work. 

\subsection{Background Equations}
\label{sec:background}
Adopting the  coordinate choice, $Q=6H^2$, the modified Friedmann equations of $f(Q)$ gravity model in the background are
\begin{eqnarray}
6H^2f_Q-\frac{f}{2}&=& \kappa^2 \rho_M 
\,, \label{eom:fQ:back:scalar:1}  
\\
2 \big(12H^2f_{QQ}+f_Q\big)\dot{H}&=&-\kappa^2(\rho_M+p_M)
\,, \label{eom:fQ:back:scalar:2} 
\end{eqnarray}
with $f_Q \equiv f_Q (Q)\equiv\partial f/\partial Q$ and $f_{QQ} \equiv f_{QQ} (Q)\equiv\partial^2 f/\partial Q^2$. The total matter density is splitted into pressureless ($p_m = 0$) non-relativistic matter density ($rho_m$) including \ac{CDM} and baryons, and relativistic radiation density, $\rho_r$, satisfies $\rho_M=\rho_m + \rho_r$. $p_m$ is the pressure of non-relativistic matter. We are imposing the condition of spatial flatness that is strongly supported by \ac{CMBR} observations \cite{Planck:2015fie}. 
Each species separately satisfies the standard continuity equations, $\dot{\rho_i}=-3H(\rho+p_i)$.

The Lagrangian with linear $Q$ is equivalent to \ac{GR}\cite{BeltranJimenez:2017tkd}, that closely follow the background of the universe. Therefore, one can think the non-GR modifications encodes through additional $F(Q)$ functions, satisfies $f(Q)=Q+F(Q)$. That does not loose the generality, because one could always defne backward $F(Q) \equiv f(Q)-Q$. In the background equations in form of explicit $F(Q)$, 
\begin{eqnarray} 
3H^2+6H^2 F_Q - \frac{F}{2}&=& \kappa^2 \rho_{\rm m} + \kappa^2 \rho_{\rm {r}}
\,, \label{eom:fQ:Friedmann:F:explicit}
\\
2\dot{H} \big(12H^2F_{QQ}+1+F_Q\big) \nonumber &&\\
+6H^2(1+F_Q)-\frac{6H^2+F}{2} &=&-\kappa^2 p_{\rm m} -\kappa^2 p_{\rm r} 
\,, \label{eq:acceleration}
\end{eqnarray}
with the derivatives with respect to $Q$, $F_Q \equiv F_Q (Q)\equiv\partial F/\partial Q$ and $F_{QQ} \equiv F_{QQ} (Q)\equiv\partial^2 F/\partial Q^2$. The above equations can be written in a standard Friedmann-like form, 
\begin{eqnarray} 
3H^2&=& \kappa^2 (\rho_{\rm m} + \rho_{\rm r}+\rho_{\rm de})
\,, \label{eq:Friedmann-eq}
\\
2\dot{H} +3H^2 &=&-\kappa^2 (p_{\rm m} + p_{\rm r}+p_{\rm de})
\,, \label{eq:acceleration-eq}
\end{eqnarray}
where the effective dark energy sector is identified as
\begin{eqnarray}
\rho_{\rm de} &\equiv& \frac{1}{\kappa^2}\left( \frac{F}{2} -6H^2 F_Q \right)\;, 
\label{eqn:rhode} \\
p_{\rm de} &\equiv& \frac{1}{\kappa^2}\left( 2 \dot{H} ( 12 H^2F_{QQ} + F_Q) - \frac{F}{2} + 6H^2 F_Q \right) \;.
\label{eqn:pde}
\end{eqnarray}

We define the usual critical density of the Universe from From Eq.~\eqref{eq:Friedmann-eq}, 
\begin{equation}
\rho_{\rm c} \equiv \frac{3H^2}{\kappa^2} \;.
\end{equation}
This allows us to introduce the fractional energy density parameters,
\begin{equation}
\Omega_{\rm m} \equiv \frac{\rho_{\rm m}}{\rho_{\rm c}} \,, \, 
\Omega_{\rm r} \equiv \frac{\rho_{\rm r}}{\rho_{\rm c}} \,,
\quad \Omega_{\rm de} \equiv \frac{\rho_{\rm de}}{\rho_{\rm c}} = \frac{F}{6H^2} - 2F_Q \;. \label{eq:Omega-parameters}
\end{equation}
The subscript $0$ denotes present-day values of these quantities. Introducing the normalized Hubble parameter, $E\equiv H/H_0$, the modified Friedmann equation Eq.~\eqref{eq:Friedmann-eq} is recasted as 
\bea 
E^2 
&=& \Omega_{{\rm m}0}(1+z)^3 
+ \Omega_{{\rm r}0}(1+z)^4
+ \Omega_{{\rm de}0} R_{\rm de}  \,
\label{eq:Friedmann-eq-E}
\eea
where the dark energy evolution factor is 
\beq
R_{\rm de} \equiv \frac{\rho_{\rm de}}{\rho_{\rm{de}0}} = \frac{ F - 12H^2 F_Q }{F_0 - 12H_0^2 F_{Q_0} } \,.
\;, \label{eq:Rde}
\eeq
The spatial flatness condition then requires
\beq
\Omega_{{\rm de}0} = 1- \Omega_{{\rm m}0} - \Omega_{{\rm r}0} = \frac{\rho_{\rm de 0}}{\rho_{\rm c 0}} =\frac{F_0}{6 H_0^2} - 2 F_{Q_0}\;. \label{eq:flatness-cond}
\eeq

The dark energy equation of state parameter, $(w_{\rm de}) ={p_{\rm de}}/{\rho_{\rm de}}$, is defined as
\bea
w_{\rm de} &=& \frac{-F +4 \left(3
   H^2+\dot{H}\right)F_Q + 48 H^2 \dot{H} F_{QQ}}{F-12 H^2 F_Q}\,.
   \label{eq:DE-EOS}    \; \label{eq:wde}
\eea

Finally, the luminosity distance $D_{\rm L}(z)$ and transverse comoving distance $D_{\rm M}(z)$ can be obtained directly from the dimensionless Hubble expansion rate, $E(z)=H(z)/H_{0}$, 
\begin{equation}
    D_{\rm L}(z) = \frac{c}{H_{0}}\int_{0}^{z}\frac{dz'}{E(z')}\, \equiv (1+z) D_{\rm M}(z)\,.,
\end{equation}
In our work, we numerically solve $E(z)$ from the aforementioned background equations. 

\subsection{Perturbation equations}
\label{sec:perturb}
We now summarize the perturbative sector of $f(Q)$ gravity model, focusing on the scalar and tensor modes relevant for large-scale structure formation and gravitational wave propagation.

\subsubsection{Scalar perturbations}
The impact of modified gravity on scalar fluctuations can be expressed through the effective gravitational coupling $\mu(a,k)$. \cite{Amendola:2007rr},
Modification {\it Effective gravitational coupling}, $\mu$,  
can be written in Fourier space  as, 

In Fourier space, the generalized Poisson equation takes the form \cite{Amendola:2007rr}
\begin{eqnarray}
\label{eq:muSigma}
&&-\frac{k^2}{a^2} \Psi = 4 \pi G_N \,\mu(a, k)  \rho_\mathrm{m} \delta_{\rm m}\,,
\end{eqnarray}
where $\Psi$ denotes the Newtonian potential, $\delta_{\rm m}=\delta\rho_{\rm m}/\rho_{\rm m}$ is the matter density contrast with respect to the background, and $k$ is the comoving wavenumber.

The {\it effective gravitational coupling} deviates from standard GR quantifies by the function $\mu$, which is the function of $F(z)$. The $\Lambda$CDM limit is $\mu\to 1$. 
Under the quasi-static approximation, which applies on sub-horizon scales, one obtains \cite{Jimenez:2019ovq},
\begin{eqnarray}
         \mu(a)&=&\frac{1}{f_Q}\, \equiv\frac{1}{1+F_Q}\,.
     \label{eq:Possion-mu}
\end{eqnarray}

It is worth noting that different conventions for the effective gravitational coupling $\kappa^2_{\rm eff}$ exist in the literature, along with corresponding definitions of the critical density $\rho_c$ and the effective dark energy density $\rho_{\rm de}$. A careful discussion of these choices and their interpretation on comparisons with $\Lambda$CDM is given in \cite{Karmakar:2025iba}.


\subsubsection{Tensor perturbations }
Tensor perturbations describe the dynamics of gravitational waves (GWs). In conformal time, the propagation equation for the tensor mode $h_{(\lambda)}$ in $f(Q)$ gravity reads
\begin{equation}\label{eom:fq:gw:conformal}
h''_{(\lambda)}+2\mathcal{H}\left(1+\frac{\diff \log f_Q}{2\mathcal{H}\diff \eta}\right) h'_{(\lambda)}+k^2h_{(\lambda)}=0.
\end{equation}

The most general GW propagation equation is given by
\cite{Nishizawa:2017nef}
\begin{equation}\label{eom:general:gw:conformal}
h''_{ij}+\left(2+\nu\right) \mathcal{H}  h'_{ij}+(c_T^2k^2 + a^2\mu^2) h_{ij}= a^2 \Gamma\gamma_{ij}.
\end{equation}
$\nu$ measures deviations from the standard amplitude damping, $c_T$ is the GW propagation speed, $\mu$ an effective mass term, and $\Gamma\gamma_{ij}$ possible source contributions.

$f(Q)$ models predict $c_T=1$, i.e. gravitational waves propagate at the speed of light, consistent with current observational constraints. The modification manifests through the damping parameter,
\begin{eqnarray}
\nu&=& \frac{1}{\mathcal{H}} \frac{\diff \log f_Q}{\diff \eta} \equiv \frac{1}{H}\frac{\diff \log f_Q}{\diff t} \,, \nonumber \\
&=& 12\dot{H}\frac{f_{QQ}}{f_Q} \;,\\
&=& 12\dot{H}\frac{F_{QQ}}{1+F_Q} \;. \label{exp:nu:FQ-FQQ-Hdot}
\label{eq:GW-nu}
\end{eqnarray}
The prime denotes differentiation with respect to conformal time $\eta$, follows relation $dt=a,d\eta$.
\section{Modeling} 
\label{sec:modeling}
In the present work, in contrast to the usual approach, where in $f(Q)$ is assumed to be a specific function of $Q$, we explore the `model-independent' approach, where we do not assume any specific form of $f(Q)$, but rather parametrize the dark energy contribution. To this end, we start by modeling the dark energy density ($R_{\rm de}$) in \cref{eq:Friedmann-eq} as a $3^{\rm rd}$ order truncated Taylor expansion\footnote{We assume that the $3^{\rm rd}$ order Taylor expansion has sufficient radius of convergence to `correctly' estimate the dark energy density in the redshift range ($a \leq 0.33$) of data available, which is validated also by our estimation of the DE EoS which is in agreement with the results presented in \cite{DESI:2025zgx}.} in the scale factor $a$ as
\begin{equation}
\label{eq:DE-density}
\frac{\rho_{\rm de}}{\rho_{\rm de0}} =  1 + \alpha_1 (a-1) + \alpha_2 (a-1)^2 + \alpha_3 (a-1)^3\,,
\end{equation}
where $\rho_{\rm de0}$ is the present value of the dark energy density, and $\alpha_i$ are the model parameters. Then $\Omega_{\rm de0} = \rho_{de0}/\rho_{c0}$, is straight away inferred from the closure condition, $\Omega_{\rm de0} = 1 - \Omega_{\rm m0}-\Omega_{\rm r0}$, where $\Omega_{\rm m,r}$ are the present day matter and radiation densities, respectively. This parametrization enusre that the dark energy density is a smootly varying function with no sharp discontinuities. Also it is not restricted to be  positive definite at all times. Once the dark energy density is obtained through the Bayesian inference, we then translate the dark energy density to the $F(Q)$ function, solving the first order ordinary differential equation following \eqeqref{eqn:rhode},
\begin{equation}
F(Q) = 6H^2 \left( \Omega_{\rm de} + 2 F_Q(Q) \right)\,,
\end{equation}
where $F_Q$ is the derivative of the function $F(Q)$. We do so for each of the sample in the posterior distribution, solving the above differential equation numerically, assuming $Q=6H^2$ as coordinate choice.In order to solve the above first order differential equation one can make assumptions regarding two initial conditions, either on the value of \(F(Q)\) itself or it's derivative $F_Q(Q)$, which imply a possibility for two different approaches, where in we can assume one to obtain the other. The two approaches then are: 
\begin{itemize}
    \item Approach-I: $ F(Q)|_{z\gg 0} \to 0$, implies the early-time is described by a GR like scenario with no cosmological constant ($\Lambda \to0$). Essentially a Einstein-de Sitter universe that evolves into an late-time accelerating phase, following the redsfhit evolution of $F(Q(z))$.
    \item Approach-II: $ F_Q(Q)|_{z\gg 0} \to 0$, implies the early-time is described by a GR like scenario with non-zero cosmetological constant ($\Lambda \to$ const) and a small derivative, allowing later redsfhit evolution. Here the cosmological constant can remain a constant or evolve. 
\end{itemize}

We remain with the approach-II as in a cosmological constant is present which can however also imply  $\Lambda = 0$, within in the posterior limits. We also test the Approach-I that tends to prouduce similar resutls however with sliglhtly larger deviations from $\Lambda$CDM, unless the inital redshift is choosen to be sufficiently large. In the analysis presented here, we assume the initial condition at $z=5 (a\sim0.2)$, which is well outside the range of observational data, alos being conservative for the radius of convergence of the taylor expansion to not fail. 



\section{Data and Analysis} \label{sec:data}
We briefly describe the observational datasets and the statistical analysis framework, which is very close as implemnted in \cite{Karmakar:2025iba, Tyagi:2025zov} and elaborated thenin. We utlise a combination of Type Ia supernovae (SNe), baryon acoustic oscillations (BAO), and early-universe priors from the CMB. 

\begin{itemize}
    \item \textit{Supernovae (SNe):}  
    We use the Pantheon+ sample of 1550 SNe Ia ($0.001 \leq z \leq 2.26$) \cite{Scolnic:2021amr} and the DESyr5 sample of 1830 SNe Ia ($0.02 \leq z \leq 1.13$) \cite{DES:2024haq}. Both are treated as uncalibrated in the joint analysis: Pan+ fits the absolute magnitude $M_b$ as a free parameter, while DESyr5 uses analytic marginalization. The likelihood is Gaussian in the distance modulus residuals,
    \[
        \mathcal{L}_{\rm SNe} \propto 
        \exp\!\left[-\tfrac{1}{2} 
        \sum_{i}\frac{(\mu_{\rm th}(z_i)-\mu_{\rm obs}(z_i))^2}{\sigma_{\mu,i}^2}\right],
    \]
    where $\mu_{\rm th}(z) = 5\log_{10}[D_L(z)/\text{Mpc}] + 25$ and $D_L(z)=(1+z)D_M(z)$.
    
    \item \textit{Baryon Acoustic Oscillations (BAO):}  
    We use DESI DR2 measurements of $D_M(z)/r_d$ and $D_H(z)/r_d$ in six bins spanning $0.1 \leq z \leq 4.2$ \cite{DESI:2025zgx}. The BAO likelihood is
    \[
        \mathcal{L}_{\rm BAO} \propto 
        \exp\!\left[-\tfrac{1}{2}\sum_{i} 
        \Delta \vec{\mu}_i^{\,T} C_i^{-1} \Delta \vec{\mu}_i\right],
    \]
    where $\Delta \vec{\mu}_i = (D_{M,i}^{\rm obs}-D_{M,i}^{\rm th}, \, D_{H,i}^{\rm obs}-D_{H,i}^{\rm th})$.
    
    \item \textit{Inverse Distance Ladder (IDL) Priors:}  
    To anchor BAO we adopt early-universe priors on the sound horizon $r_d$ and recombination Hubble parameter $H_{\rm rec}$ from Planck 2018, with covariance from \cite{Haridasu:2020pms}(equivalent to the approach in \cite{Lemos:2023xhs,DESI:2024mwx,DESI:2025fii,Lynch:2025ine}). The likelihood is Gaussian,
    \[
        \chi^2_{\rm IDL} = 
        \Delta \boldsymbol{\mu}_{\rm IDL}^{T}\, C^{-1}\, \Delta \boldsymbol{\mu}_{\rm IDL},
    \]
    with $\Delta \boldsymbol{\mu}_{\rm IDL}=\{\Delta H_{\rm rec},\Delta r_d\}$.
\end{itemize}
These datasets are complementary in redshift coverage and precision, and together provide a robust probe of the expansion history. The joint likelihood is
\begin{equation}
    \ln \mathcal{L} = \ln \mathcal{L}_{\rm SNe} + \ln \mathcal{L}_{\rm BAO} + \ln \mathcal{L}_{\rm IDL}.
\end{equation}

The posterior sampling is performed with \texttt{Nautilus} \cite{Lange:2023ydq}, and ingerence of constraints is performed using \texttt{GetDist} \cite{Lewis:2019xzd}.

\begin{figure*}[htbp]
\centering
\includegraphics[scale = 0.35]{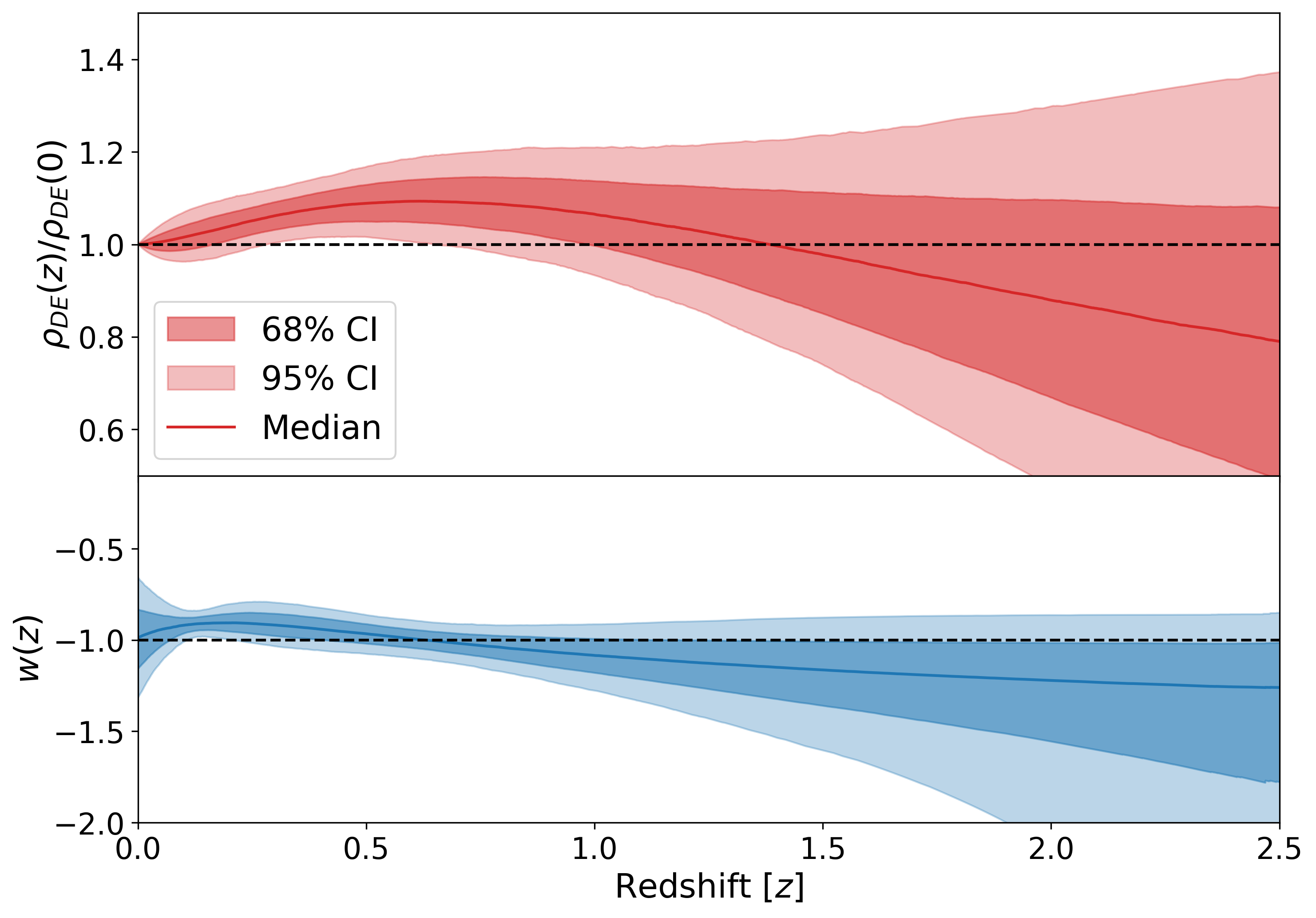}
\includegraphics[scale = 0.35]{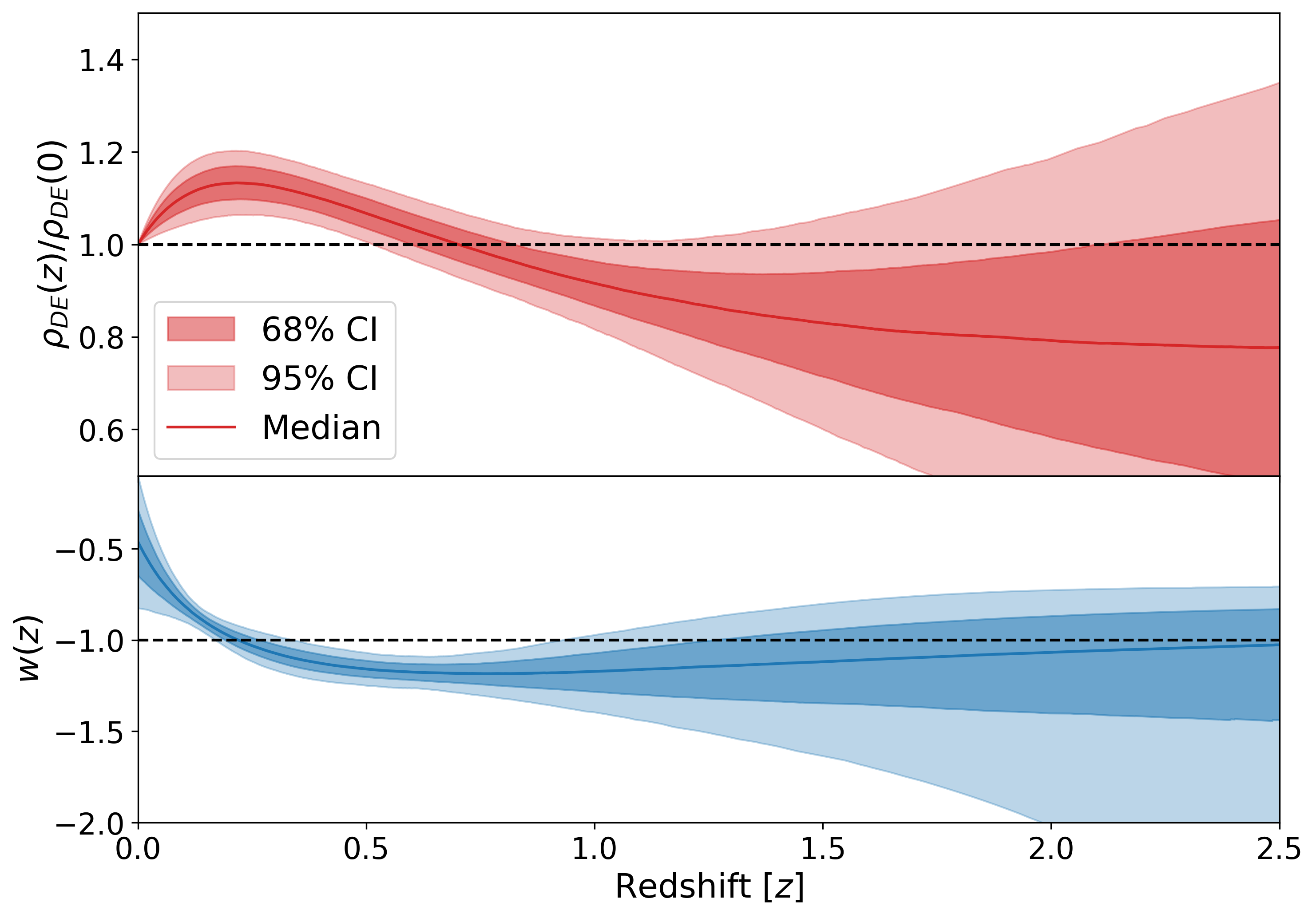}
\caption{Posterior distributions of the dark energy density (\textit{top}) as function modeled as a 3rd order Taylor expansion in the scale factor $a$ as described in \eqeqref{eq:DE-density} and the corresponding DE equation of state (\textit{bottom}) parameter $w_{\rm de}$, defined in \eqeqref{eq:DE-EOS}, as a function of the redsfhit. The dashed line corresponds to the cosmological constant case, $w_{\rm de} = -1$. The \textit{left} and \textit{right} panels correspond to the dataset combination DESI+\Panp and DESI+DESyr5, respectively.}
\label{fig:DE-density}
\end{figure*} 

\section{Results and Discussion} 
\label{sec:resutls}
 
We begin by firstly presenting the posterior distributions of the dark energy density from the joint analysis of SNe and BAO datasets. In \Cref{fig:DE-density}, we show the posterior distribution of the reconstructed dark energy density (top panel) and the corresponding equation of state parameter $w_{\rm de}$ (bottom panel) as a function of the redshift. The left and right panels correspond to the dataset combination DESI+\Panp and DESI+DESyr5, respectively. The dashed line in the top panel corresponds to the cosmological constant case, $\rho_{\rm de} = \rho_{\rm de0}$, which accordingly in the bottom panel corresponds to $w_{\rm de} = -1$. The shaded region are the $1\sigma$ and $2\sigma$ C.L. of the reconstructed dark energy density and equation of state parameter $w_{\rm de}$. 

We note that there is a significant difference between the dark energy density reconstructed from the two dataset combinations driven by the different SNe datasets as expected \cite{DESI:2024mwx,DESI:2025zgx, DESI:2025brt}. The \Panp dataset shows a mild deviation from the cosmological constant case, $\rho_{\rm de} = \rho_{\rm de0}$, at lower redshift, $z \lesssim 1.5$, while the DESyr5 dataset shows a larger deviation as is also reported in \cite{DES:2024haq}. This difference is also reflected in the posterior distribution of the equation of state parameter $w_{\rm de}$, where the DESyr5 dataset shows a larger deviation from the cosmological constant case, $w_{\rm de} = -1$, at lower redshift, $z \lesssim 1.5$. Also corroborating the detection of the phantom crossing, which has received a overwhelming attention. {This difference could possibly be attributed to the different systematics in the two datasets, which are yet to be strictly assessed in the relatively newer DESyr5 compilation \cite{Efstathiou:2024xcq}.}

As can be seen in \cref{fig:DE-density}, the posterior distribution of the reconstructed dark energy density is very well constrained with deviations of the order of $1-2\sigma$ from the cosmological constant case, $w_{\rm de} = -1$, when using the DESI and \Panp dataset. Note that when parameterizing the DE EoS instead of the DE density as an higher-order expansion is expected to reduce the significance of the detection. {This is very well understood and expected in the higher order expansions of the EoS, see for instance \cite{Nesseris:2025lke}} at higher redsfhit and shows reduced evidence for a phantom crossing as expected from the standard CPL parametrization, as reported in \cite{DESI:2025brt,DESI:2025ejh}. However, in our analysis, we parametrize the DE density, which is a more direct constituent of the expansion rate, and the EoS itself is a later translation of background quantities. And as expected we find a strong evidence for a phantom crossing, $w_{\rm de} < -1$, at lower redshift, $z \lesssim 1.5$, in agreement with the results presented in \cite{DESI:2025zgx,DESI:2025brt}, when utilizing the DESI+DESyr5 dataset combination.

\subsection{Reconstruction of the $f(Q)$ function and modified gravity parameters}

\begin{figure*}[htbp] 
\centering
\includegraphics[scale = 0.29]{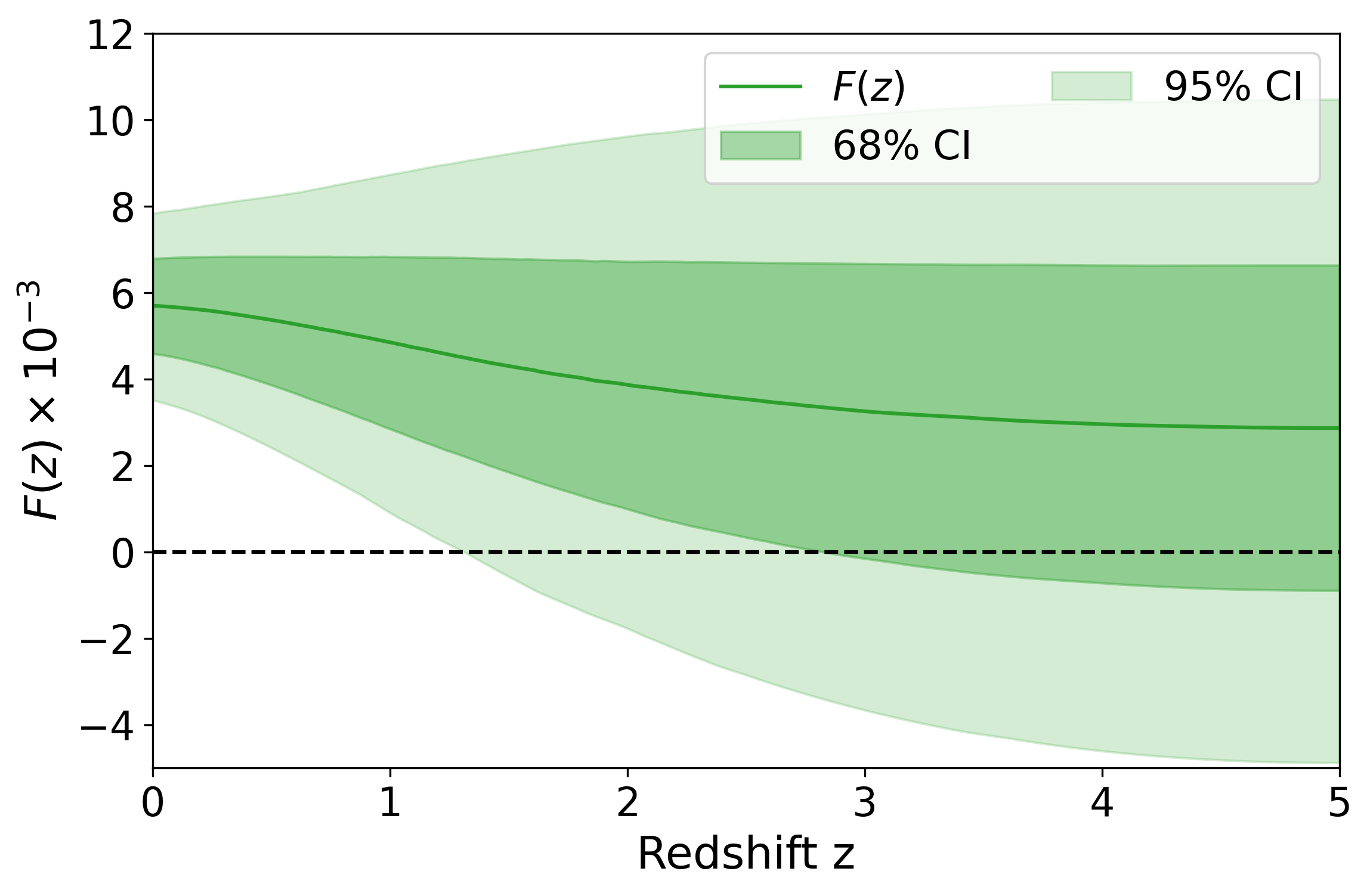}
\includegraphics[scale = 0.29]{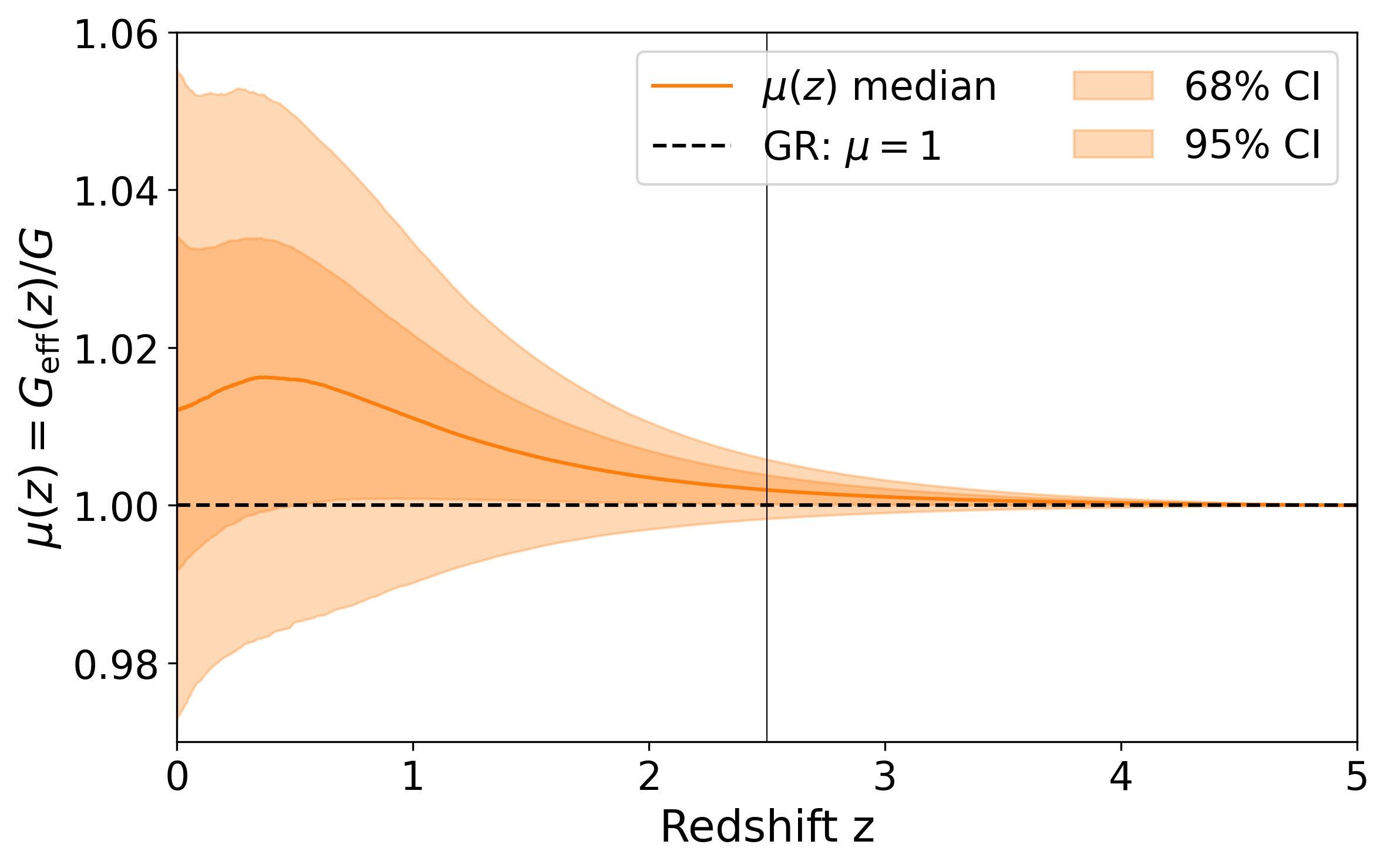}
\includegraphics[scale = 0.29]{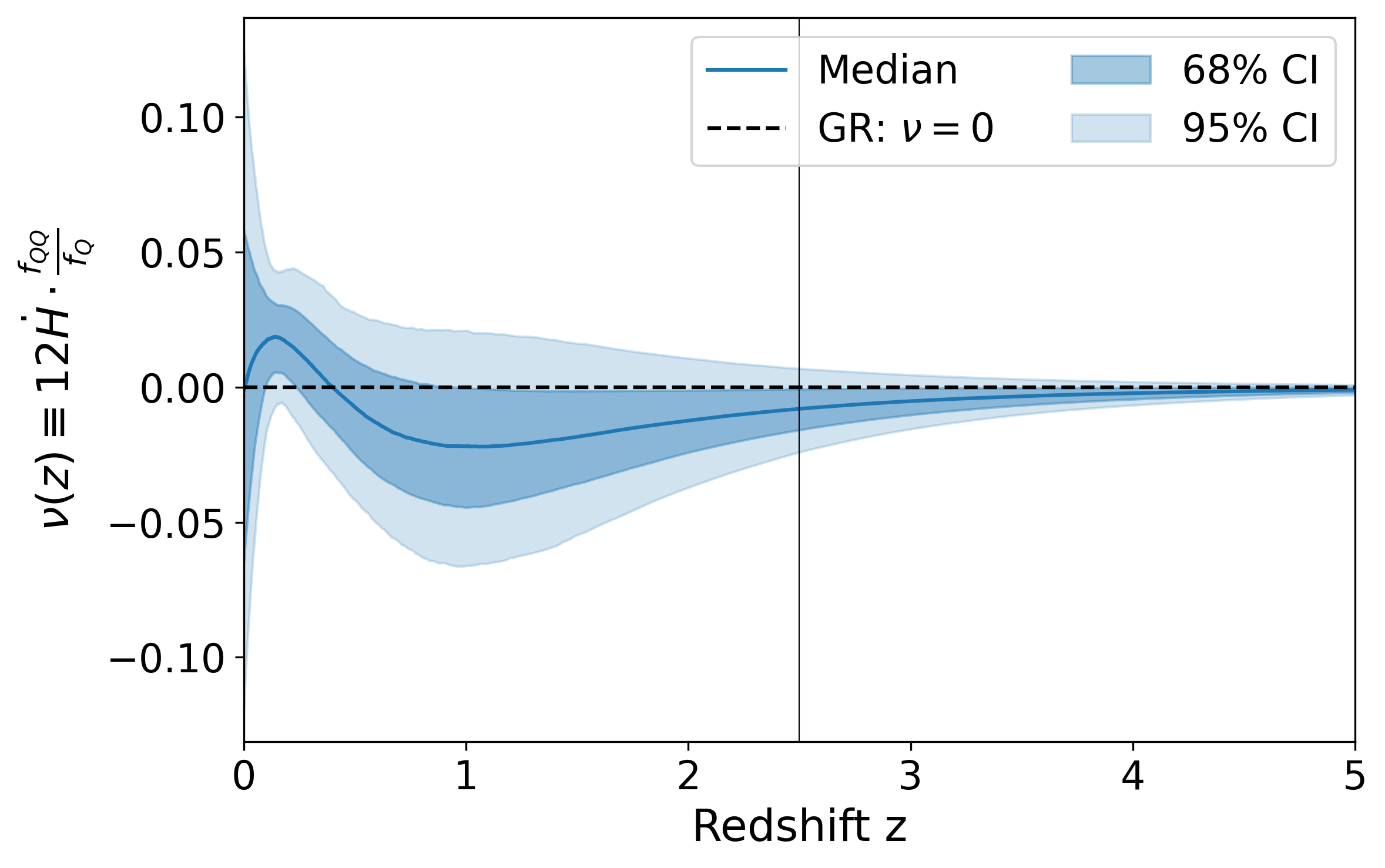}
\includegraphics[scale = 0.29]{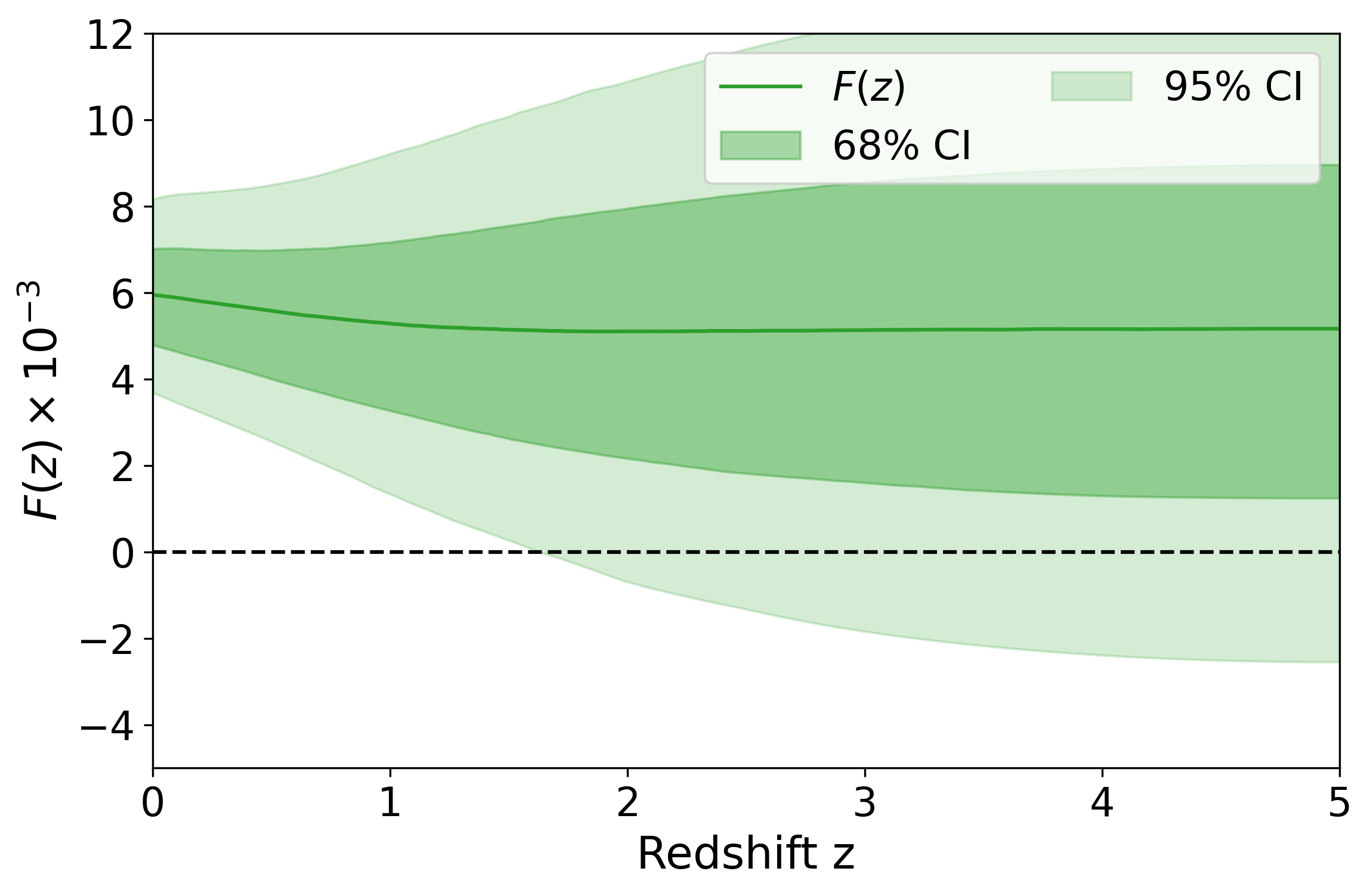}
\includegraphics[scale = 0.29]{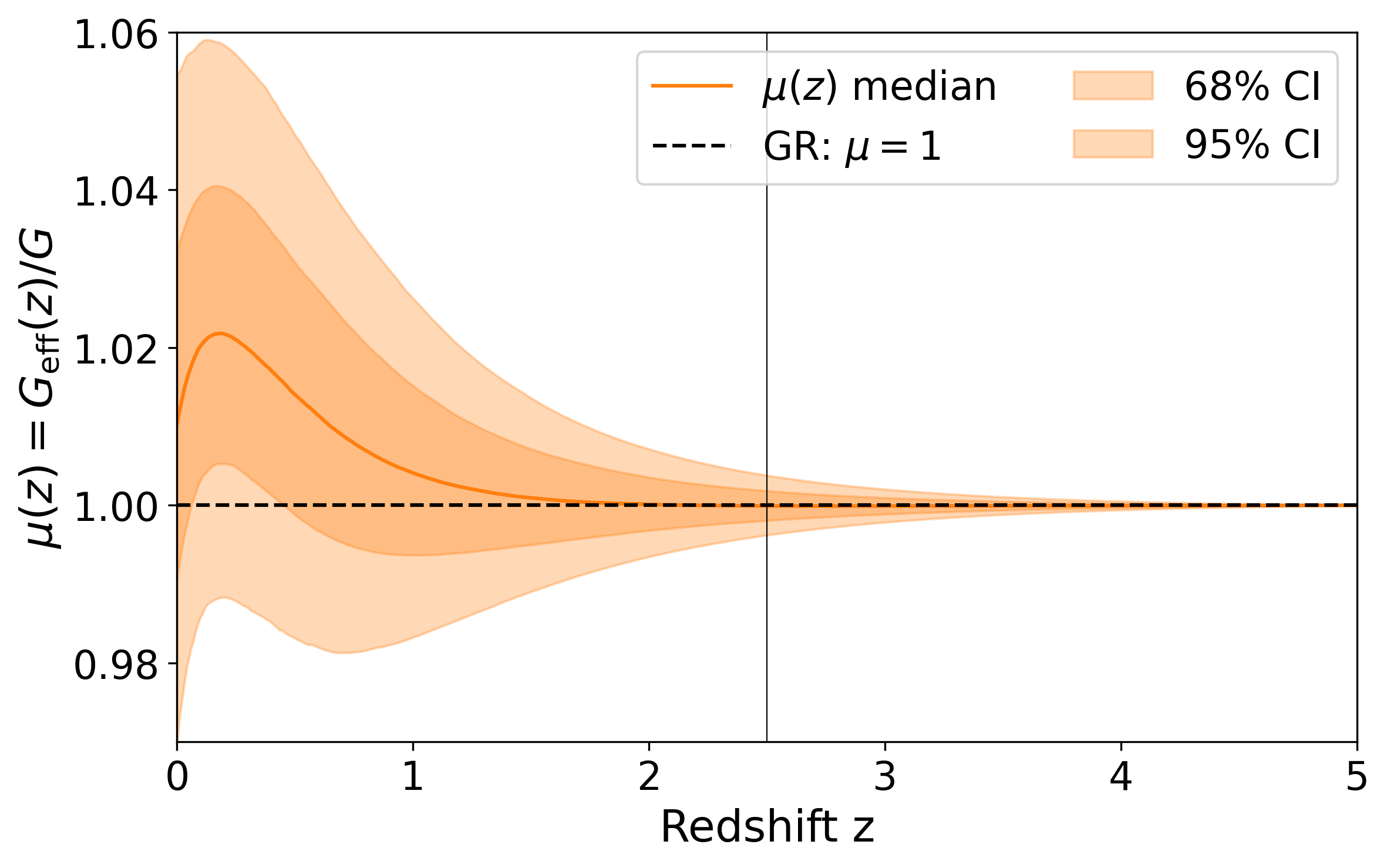}
\includegraphics[scale = 0.29]{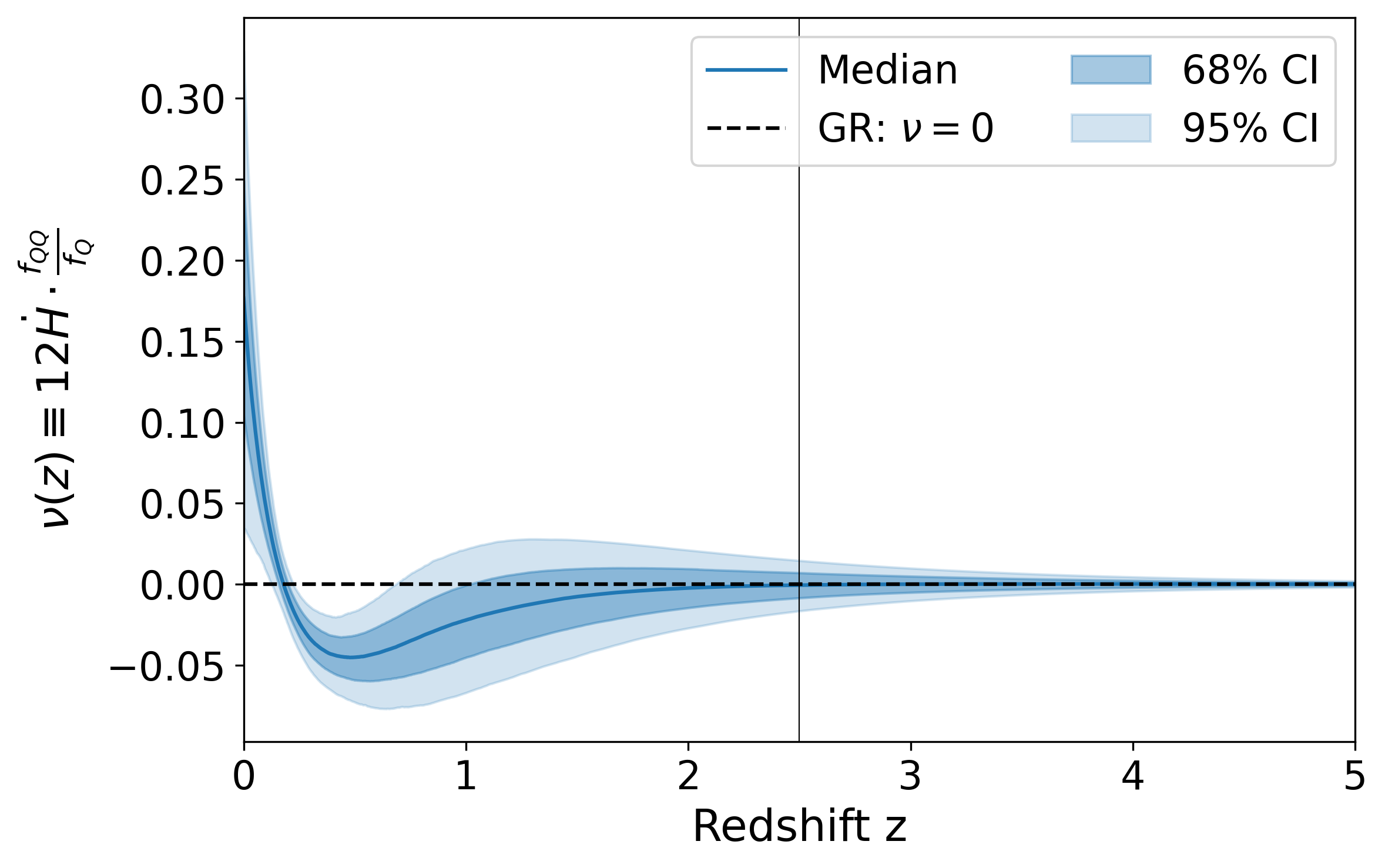}

\caption{\textit{Top} row corresponds to the data combination DESI+\Panp and the \textit{bottom} corresponds to the DESI+DESyr5. We show the posterior distribution of $F(z)$ (\textit{left}), the effective gravitational coupling $\mu$ (\textit{center}) and the damping factor $\nu$ (\textit{right}) as a function of the redshift. The dashed line in the latter two corresponds to the cosmological constant case, $\mu = 1,\, \nu = 0$. The shaded region corresponds to the $1\sigma$ and $2\sigma$ C.L. of the reconstructed quantities. }
\label{fig:STG-reconstruction}
\end{figure*}

Having estimated the 'model-independent' evolution dark energy, we now turn to the reconstruction of the $f(Q)$ function from the dark energy density, following the approach described in \Cref{sec:modeling}. The posterior distribution of the $F(Q)$ as function of redsfhit is shown in \cref{fig:STG-reconstruction}. We find each of these quantities well-constrained essentially implying a very good correspondence between the reconstructed dark energy density and the unassumed $f(Q)$ function. 

The reconstructed $F(z)$ function is consistent with the cosmological constant case, $F(z) = 6H^2 \equiv Q$, within $1\sigma$ level all redshifts. While at the higher-redshifts the posterior distribution of $F(z)$ is also very well consistent with the case, $F(z) = 0 $ ($F_0$), at $\sim 1\sigma$ level, for both the dataset combinations DESI+\Panp and DESI+DESyr5. However, at lower redshift, $z \lesssim 1.5$, the posterior distribution of $F(z)$ shows a strong deviation from this $F(z) = 0 $ case, indicating the presence of at least a cosmological constant like term, ensuring late-time acceleration. The deviation is more pronounced for the DESI+DESyr5 dataset combination, which is also reflected in the reconstructed dark energy density and equation of state parameter $w_{\rm de}$ as shown in \cref{fig:DE-density}. We find a constraint of $F_0/10^{3}= 5.95^{+1.15}_{-1.13} $ and asymptotic high-redshift value of $F_{z\gg0}/10^{3}= 5.20^{+4.05}_{-3.83} $, for the dataset combination DESI+DESyr5. Similarly for the DESI+\Panp combination we find $F_0/10^{3}= 5.64^{+1.15}_{-1.08} $ and $F_{z\gg0}/10^{3}= 2.73^{+3.95}_{-3.75} $, showing a lower value of the latter wrt the DESI+DESyr5 dataset.

\textit{Scalar perturbations:} Then the reconstruction of running of the gravitational constant $\mu = 1/(1+F_Q)$\footnote{The posterior distribution of $F_Q$ function is mostly consistent with the, $F_Q \sim 0 $, within $1\sigma$ level at higher redshift and the lower $z \lesssim 1.5$. The bottom panel of \cref{fig:STG-reconstruction} shows the full $F(z)$ which deviates significantly at lower redshift, $z \lesssim 1.5$ indicating an acceleration, while slowly tending to $F(z>3) \to 0$ at higher redshifts. } can be inferred from the reconstructed $F_Q$ function. The posterior of the effective gravitational coupling $\mu$ is shown in the center-panel of \cref{fig:STG-reconstruction}. The posterior distribution of $\mu$ is consistent with the cosmological constant case, $\mu = 1$, well within $1\sigma$ level at higher redshift and shows a deviation at lower redshift, $z \lesssim 1.5$. We find a $1\sigma$ deviation locally, $z =0$, indicating a deviation from the standard $\Lambda$CDM model. Our reconstruction indicates a mild increase of the effective gravitational coupling reaching $\mu \sim 1.01 \pm 0.02$ which is a $2\%$ increase wrt the Newtonian $\GN$. Interestingly, this inference is completely consistent between the two datasets combinations DESI+\Panp and DESI+DESyr5.

This finding also implies that the $\sim 2\%$ variation in the gravitational constant can yield very little advantage in addressing the $H_0$ and $S_{8}$-tensions (see \cite{CosmoVerseNetwork:2025alb} for a review). Our results can be contrasted against the recent findings in \cite{Boiza:2025xpn}, where the implications of  $f(Q)$ gravity for the tensions was discussed assuming specific parametric model of $F(Q)$ using the similar dataset combinations. 

\textit{Tensor perturbations:} Finally, we reconstruct the damping parameter $\nu$ from the reconstructed $F_Q$ and $F_{QQ}$ functions using \eqeqref{eq:GW-nu}. The reconstructed running parameter $\nu$ is shown in the right-panel of \cref{fig:STG-reconstruction}. The posterior distribution of $\nu$ is mostly consistent with the cosmological constant case, $\nu = 0$, within $1\sigma$ level within $z\lesssim 2.5$ where data is available for the DESI+\Panp combination. However, we find the median curve of $\nu(z)$ to be negative, $\nu \sim -0.02 $ at $z\sim 1$ reminiscent of the predictions made in a exponential model fro $F(Q$) or the inverse logarithmic model we introduced in \cite{Karmakar:2025iba}, which then transitions to positive values at $z\sim 0.2$. We find a constraint of $\nu= 0\pm 0.06$ after the said oscillations in redshift.

Now turning to the most crucial finding of the current analysis, the reconstructed $\nu$ using the DESI+DESyr5 dataset shows a strong signature for a deviation from GR like scenario. This signature, interestingly remains similar in nature to that from DESI+\Panp, however, more pronounced. As shown in the right-bottom panel of the \cref{fig:STG-reconstruction}, we find a steep increase in the redshift range $0\lesssim z \lesssim0.5$ reaching from $\nu \sim -0.05$ at $z\sim 0.5$ to $\nu\sim 0.15$ at $z=0$. This indicates a strong deviation from the GR like scenario, $\nu = 0$, at $>2\sigma$ level locally, $z\sim0$. This finding is also consistent with the strong deviation from the cosmological constant case, $w_{\rm de} = -1$, as shown in \cref{fig:DE-density}. We find a constraint of $\nu = 0.18^{+0.08}_{-0.07}$ at $68\%$ C.L., which is a $\gtrsim 2\sigma$ deviation from the GR scenario, $\nu = 0$. We stress once again, that in our current analysis, we find that the DESI+\Panp dataset combination does not show such a strong deviation from GR like scenario, $\nu = 0$, which could be attributed to the different systematics in the two SNe datasets as discussed earlier. 

Note that most traditional parametric models do no allow for a transition of the $\nu$ which remain either positive or negative throughout the cosmic history. At a face-valve, one can immediately infer that that the nature of the $f(Q)$ function is such that it allows for a transition of the running parameter $\nu$ at lower redshift and hence demonstrates the inadequacy of several proposed models and subsequent tests performed in literature. On the other hand, it remains an important prediction for the gravitational wave propagation that can be crucial to distinguish between different modified gravity models and more importantly form dark energy models within GR. 

In the context of phenomenological hints/evidence for dynamical dark energy, such a behavior sourced by fundamental modified gravity model will display varied predictions in the scalar and tensor sectors. The current analysis indicates that the DESI+DESyr5 dataset combination shows a strong evidence for a dynamical dark energy, which is also reflected in the reconstructed $f(Q)$ function and the modified gravity parameters $\mu$ and $\nu$.

\section{Conclusions} 
\label{sec:conclusions}
In the work, we have performed a 'model-independent' reconstruction of the $f(Q)$ cosmology, as an example, from the latest SNe and BAO data. The analysis has been performed to assess the implications for the modified gravity theories in the context of the recently reported hints for a dynamical dark energy. We have used two different SNe datasets, the \Panp and the DESyr5, in combination with the latest BAO measurements from DESI-DR2.

We find that the reconstructed $F(Q)$ function is consistent with the cosmological constant case, $f(Q) = Q + \Lambda$, in agreement with $\Lambda = 0$ within $1\sigma$ level at high-z. While showing a significant evidence at lower redshifts, $z \lesssim 1.5$ being different from $F_0\equiv \Lambda =0$, necessary to source the late-time acceleration. We then estimated the deviation of effective gravitational coupling $\mu$ and the damping parameter of GW propagation $\nu$ from the reconstructed $f(Q)$ function. We find that $\mu$ is consistent with $\mu = 1$, within $1\sigma$ level, while showing a mild increase at lower redshifts, reaching $\mu \sim 1.02 \pm 0.02$ at present time. 

The gravitational wave amplitude damping parameter $\nu$ is mostly consistent with standard scenario, $\nu = 0$, within $1\sigma$ level within $z\lesssim 2.5$, but shows a negative median value at higher redshifts, reminiscent of predictions made in inverse logarithmic model of $f(Q)$ using the DESI+\Panp dataset \cite{Karmakar:2025iba}. However, the DESI+DESyr5 dataset combination shows a strong deviation from the GR like scenario, $\nu = 0$, at $>2\sigma$ level locally, $z=0$, reaching $\nu = 0.18^{+0.08}_{-0.07}$ at $68\%$ C.L. This finding is also consistent with the strong deviation from the cosmological constant case, $w_{\rm de} = -1$, as shown in \cref{fig:DE-density}. This in fact implies a possibility for future gravitational wave standard siren measurements to further constrain the $\nu$ parameter and hence the $f(Q)$ function, which can distinguish between different modified gravity models and dark energy models within GR, proving equivalent dynamical dark energy behavior. 

Our results on the one hand provides necessary insights into the nature of the $f(Q)$ function and its implications for modified gravity theories in the context of model building going beyond the simple parametric forms explored so far in literature. On the other hand, we find that the results are dataset dependent, also highlights the importance of understanding and mitigating systematics in SNe datasets, as different compilations can lead to varying conclusions. We have explored a framework, that can be extended to other modified gravity models to test the phenomenological implications of the hints for a dynamical dark energy, through precise perturbation and gravitational wave data to arrive in the near future. Surveys with improved precision and control over systematics will be crucial to further elucidate the nature of dark energy and its connection to modified gravity theories.

\section*{Acknowledgments}
We would like to thank L. Järv, T. Koivisto, A. Nishizawa and M. Saal for useful discussions and comments on a preliminary version of this paper. PK acknowledged the courtesy provision of computing resources by the University of Tartu. SH is supported by the INFN INDARK grant and acknowledges support from the COSMOS project of the Italian Space Agency (cosmosnet.it).

\bibliography{bibliography} 

\end{document}